\documentstyle[12pt]{article}
\input{psbox.tex}

\def\be{\begin{equation}}
\def\ee{\end{equation}}
\def\bea{\begin{eqnarray}}
\def\eea{\end{eqnarray}}

\def\br{\begin{thebibliography}{abc}}
\def\er{\end{thebibliography}}

\def\Hat#1{\rlap{\kern.10em$\widehat{\phantom G}$}#1}
\def\HAt#1{\rlap{\kern.05em$\widehat{\phantom G}$}#1}

\def\czp#1{\rlap{\kern.1em$\widehat{\phantom{G\vrule height.8em}}$}#1{}}
\def\Czp#1{\rlap{\kern.05em$\widehat{\phantom{G\vrule height.8em}}$}#1{}}

\newcommand{\sect}[1]{\setcounter{equation}{0}\section{#1}}

\def\sxn#1{\bigskip\medskip \sect{#1} \smallskip
                                                 }

\begin{document}

\thispagestyle{empty}
\setcounter{page}{0}

\begin{flushright}
SU-4240-579\\
May 1994
\end{flushright}
\vspace*{15mm}
\centerline {\LARGE DISCRETE TIME FROM QUANTUM PHYSICS}
\vspace*{15mm}
\centerline {\large A. P. Balachandran and L. Chandar}

\vspace*{5mm}
\centerline {\it Department of Physics, Syracuse University,}
\centerline {\it Syracuse, NY 13244-1130, U.S.A.}
\vspace*{25mm}
%\baselinestretch{2.0}
\normalsize
\centerline {\bf Abstract}
\vspace*{5mm}
't Hooft has recently developed a discretisation of (2+1) gravity which has a
multiple-valued Hamiltonian and which therefore admits quantum time evolution
only
in discrete steps.  In this paper, we describe several models in the continuum
with single-valued equations of motion in classical physics, but with
multiple-valued Hamiltonians.  Their time displacements in quantum theory are
therefore obliged to be discrete.  Classical models on smooth spatial manifolds
are also constructed with the property that spatial displacements can be
implemented only in discrete steps in quantum theory.  All these models show
that quantization can profoundly affect classical topology.

\newpage

\baselineskip=24pt
\setcounter{page}{1}

\sxn{Introduction}

In prevailing models in fundamental physics, it is generally the case that time
and space are treated as manifolds, with no fundamental limitations on their
smallest units.  The standard and grand unified models are based on Minkowski
spacetimes while in more ambitious approaches like those using strings as
well, there is no indication of any discrete structure associated with space or
time.  Although there is an occasional theorist proposing a more radical
hypothesis \cite{1}, it is nevertheless the case that such views are not
seriously entertained in the dominant ideology.

There is however no known basic principle requiring space or time to be
continuous or forbidding limitations on their units.  Quantum theory for
example does not exclude their discreteness as shown by the consistency of
lattice gauge theories \cite{2} or numerous models like the tight binding
\cite{3}, Ising \cite{4} and Hubbard models \cite{5} in condensed matter
physics and elsewhere.  Ultimately, of course, the nature of spacetime can be
understood better only by a combination of theory and practice and is not
capable of resolution merely by abstract thought.

In recent articles, 't Hooft \cite{6} has developed a scheme for discretisation
of canonical 2+1 gravity.  The Hamiltonian in his approach is proportional to
an angle when there is a point source, and is not then single-valued.  It is
only the evolution operator causing a shift in time by a basic unit $T_{0}$ and
the discrete group it generates which are well-defined and meaningful.  There
is thus a smallest unit for possible time shifts in this model of gravity.  It
is striking that this unit appears to naturally emerge from the lattice
formulation of a canonical gravity model.

In this work, we further explore the possibility of a minimum time unit and
\mbox{examine} similar limitations on spatial shifts as well.  There are
numerous
classical models on smooth spacetimes which can be quantized only if time
shifts are discrete, this being so even without approximating the spatial
manifold by a lattice.  There are also models of the same kind which admit only
discrete spatial shifts in quantum theory.  Typically the configuration spaces
of these models are multiply connected and these effects are brought about by
certain
topological terms in their Lagrangians.  The terms responsible for
discretisation
of time evolution are not unlike the topological term in magnetic monopole
theory or the Skyrme model \cite{7}, the principle difference being that such a
term
is
associated with the second homology group of the configuration space in the
latter \cite{7} and with the first homology group here \cite{8}.  The classical
equations of motion in all these cases make perfect physical sense without any
sort of discretisation.  An important point meriting emphasis is that whereas
time or
space \underline{translations} are constrained to be discrete here on
quantization, spacetime itself is allowed to be a continuum.  This
circumstance raises certain conceptual issues which will also be addressed in
this paper.

It may be noted that we can find models where time or space translations or
both are discrete in quantum theory.  There is no necessary correlation between
these properties.  Our Lagrangians of course also lack
Poincar$\acute{\mbox{e}}$ or Lorentz invariance.

Section 2 begins the discussion by studying a simple model of a particle on a
circle $S^{1}$.  The classical equations of motion are single-valued and
well-defined, but their time independent Hamiltonian is multiple-valued.
Hamiltonian is not crucial for classical physics except in statistical
mechanics which anyway is close to quantum theory.  It is not therefore a
serious matter to put up with a multiple-valued Hamiltonian in classical
physics.
But that is not the case in quantum theory.  As we argue in detail, continuous
quantum time evolution with such a Hamiltonian leads to obstructions in the
identification of $S^{1}$ as the configuration space.  If we insist on the
preservation of $S^{1}$ as the configuration space, time evolution is possible
only by discrete multiples of a basic unit $T_{0}$.  In this way, quantization
can deeply affect the nature of time in quantum theory.  It also affects energy
in consequence, which becomes an angular coordinate on a circle of radius
$\frac{2\pi}{T_{0}}$.  It is thus conserved only modulo $\frac{2\pi}{T_{0}}$.

Now there is nonuniqueness in the choice of the Hamiltonian for this dynamical
system.  Among others, there is in particular also a time-dependent choice as
we show in Section 3.  Indeed this system has a simple classical
interpretation as a charged particle on $S^{1}$ subject to a constant electric
field, and the latter can be described by electromagnetic potentials with or
without
time dependence.  The time dependent Hamiltonian is single-valued and does not
lead to quantized time evolution, but it fails to give a conserved
energy.  Thus quantum physics for these potentials with and without time
dependence are quite different although they are apparently connected by a
gauge transformation.  We discuss the reason for this inequivalence as well in
this section.  It shows clearly that discretisation of quantum time evolution
is generic for $S^{1}$ in the sense that it can be achieved for any Lagrangian
dynamics on $S^{1}$ by modifying the Lagrangian without affecting classical
evolution.

Section 4 identifies the topological basis for the existence of multiple-valued
Hamiltonians with single-valued equations of motion.  In accordance with a
previous remark, such dynamics is possible whenever the configuration space
admits a closed but inexact one form \cite{8}.  Section 4 shows the method of
construction of a multiple-valued Hamiltonian from \underline{any}
single-valued
Hamiltonian without affecting classical physics once this condition is met.
The former as usual leads to discrete
time evolution in units of a constant $T_{0}$ while the latter instead has
continuous time evolution.  Thus these quantum systems are inequivalent just as
before despite their classical equivalence.

Field theoretic examples of multiple-valued Hamiltonians are described in
Section
5.  There are many such models.  For example, the ``$O(N+1)$ non-linear
$\sigma$-models'' in $(N+1)$- dimensional spacetimes and $CP^{N}$ \cite{7,9}
models in (1+1)-
dimensional spacetimes are shown to admit such Hamiltonians.  The case $N=1$
of these
models is considered in a little detail.

There exist also quantum models where spatial translations are possible only in
quantized units as we show in Section 6.  Their classical equations of course
are once again well-defined.  Although these models too require the topological
condition mentioned above, they are otherwise different from those with
quantized time evolution.

Section 7 is the last one.  There we argue that physical consistency requires
quantized time if time evolution is quantized.  But we can find no such
argument for spatial quantization.

We have done exploratory work on the experimental limitations on time and space
quantization \cite{10}.  It suggests that these limits are quite poor.  For
example, it seems that $\frac{2\pi}{T_{0}}$ can be as small as a few
TeV without violating available data.  We hope to describe this work elsewhere.

We conclude this Section with a small observation.  Let $S_{i} (i=1,2,\ldots )$
be systems with time units $T_{i}$ for quantum evolution.  Let us allow these
systems to interact and form a composite system $S$.  If $T$ is the time
unit for $S$, then evidently $\frac{T}{T_{i}} =n_{i}$ must be integers.
This is possible only if $T_{i}/T_{j}$ are all rational so that they are
integral multiples of a common $\tau$: $T_{i}=n_{i}\tau \; \; n_{i}\in Z^{+}$.
There is no quantum time evolution for $S$ if the contrary is the case.

\sxn{Particle on a Circle}

Let $\theta$ (mod $2\pi$) be the angular coordinate on a circle $S^{1}$.  We
consider a constantly accelerating particle of mass $m$ on this configuration
space:
\be
m\ddot{\theta}\equiv m\frac{d^{2}\theta}{dt^{2}} =
-c, \;\; c= \mbox{ a constant . }  \label{2.1}
\ee

We assume hereafter that $c>0$.  The $c<0$ case becomes this one if $\theta$ is
redefined according to $\theta \rightarrow -\theta$.

We want to explore the nature of quantum physics of this dynamical system.  But
let us first underline that there is nothing pathological or exotic about
(\ref{2.1}) which is just a single-valued equation on $S^{1}$.  Furthermore, in
classical physics,
all enquiries about this system are answerable using (\ref{2.1}) except in
statistical mechanics.  The latter anyway contains  quantum principles in one
form or another and will be ignored here.

Now a familiar approach to quantization starts from a Lagrangian and subjects
it to canonical quantization.  Let us use this method for (\ref{2.1}) as well.

A Lagrangian for (\ref{2.1}) is
\be
L=\frac{1}{2}m\dot{\theta}^{2} -c\theta ,\;\;\; \dot{\theta}\equiv
\frac{d\theta}{dt}. \label{2.2}
\ee
It is a remarkable expression, being multiple-valued in $\theta$.

The Hamiltonian for (\ref{2.2}) after quantisation is
\be
H=\frac{p^{2}}{2m} +c\theta ,\;\;\; p=m\dot{\theta} \label{2.3}
\ee
where $\theta$ and $p$ have the commutators
\be
[\theta ,\theta ] = [p,p] =0,\;\;\; [\theta ,p]=i. \label{2.4}
\ee
The Schr$\ddot{\mbox{o}}$dinger equation is hence
\be
H\Psi = [-\frac{1}{2m^{2}}\frac{\partial ^{2}}{\partial \theta ^{2}}-c\theta
]\Psi =i\frac{\partial \Psi}{\partial t} \label{2.5}
\ee
while states of fixed energy are given by
\be
H\chi =E\chi . \label{2.6}
\ee

The solution of (\ref{2.5}) or (\ref{2.6}) involves the choice of boundary
conditons (BC's) for $\Psi$ and $\chi$.  The vectors fulfilling the BC's (and
suitable differentiability properties)
constitute the domain $D(H)$ of the Hamiltonian in the Hilbert space \cite{11}
\be
{\cal H} = \{ \Psi |(\Psi ,\Psi ):=\int _{0}^{2\pi}d\theta\, |\Psi
(\theta)|^{2}<\infty \} .\label{2.7}
\ee

There are technical requirements on $D(H)$ which will not be addressed here,
but
there are also physical principles limiting its choice \cite{12}.  Let us now
briefly describe them.

The space ${\cal H}$ consists of all square integrable functions.  There is no
need for continuity either of its elements or of their probability densities.
All infinite dimensional (separable) Hilbert spaces are in fact unitarily
isomorphic.  [Let us choose the orthonormal bases \mbox{\{ $\phi _{n}^{(i)}$\}}
$(n=0,1,2,\ldots ;i=1,2)$ for the two infinite-dimensional Hilbert spaces
${\cal H}_{n}^{(i)}$.  Then such an isomorphism is given by the map $\phi
_{n}^{(1)}\rightarrow \phi _{n}^{(2)}$.]  It is therefore impossible to
identify the underlying configuration space just by
examining all wave functions.  It follows that if we are also quite free to
choose $D(H)$, the probability densities of vectors in $D(H)$ may fail to be
continuous on $S^{1}$ and may be unsuitable for work on $S^{1}$.

Topology therefore enters the choice of $D(H)$ \cite{12}.  It does not quite
require that $D(H)$ consists of continuous functions $\Psi$ on $S^{1}$ as only
probability densities $\Psi ^{*}\Psi$ are observable.  So it is enough to
require that $\Psi
^{*}\Psi$ is continuous on $S^{1}$.  Since $(\Psi +\chi )^{*}(\Psi +\chi )$ and
$(\Psi +i\chi )^{*}(\Psi +i\chi )$ must also be continuous for $\Psi ,\chi \in
D(H)$, we can conclude that
\begin{eqnarray}
&&\Psi ^{*}\chi \in {\cal C}(S^{1}) \mbox{ if } \Psi ,\chi \in
D(H),\nonumber\\
&&{\cal C}(S^{1}) =\mbox{ The space of continuous functions on
}S^{1}.\label{2.8}
\end{eqnarray}

The domain fulfilling this principle is not unique.  It depends on a point
$e^{i\phi }$ of $S^{1}$ and has the definition
\be
D_{\phi}(H)=\{ \Psi |\Psi (2\pi )=e^{i\phi}\Psi (0);\; \Psi '(2\pi
)=e^{i\phi}\Psi '(0)\} ,\label{2.9}
\ee
the condition on $\Psi '$ being necessary because the differential operator $H$
is of second order.  It is to be understood also that $\Psi$ is a
suitably smooth function in the interior $]0,2\pi [$.

We recall that there is a well-known interpretation of $\phi$ in terms of a
magnetic flux threading the circle.  Notice also that quantum theory depends on
$D_{\phi}(H)$ and is not therefore unique.

The use of (\ref{2.9}) does not resolve problems of interpretation because $H$
is multiple-valued.  Thus suppose that $\Psi _{\theta _{0}}$ is a wave
function
in $D_{\phi}(H)$ peaked sharply at a point of $S^{1}$ with coordinate
$\theta _{0}$ and fulfilling also
$\Psi _{\theta _{0}}''(2\pi )=e^{i\phi}\Psi
_{\theta _{0}}''(0)$.  Physics then requires that the expectation value $(\Psi
_{\theta _{0}},H\Psi _{\theta _{0}})$ has period $2\pi$ in $\theta _{0}$ if
$H$ is a decent observable.  But that clearly is not the case, $H$ being
multiple-valued.  We can not therefore accept $H$ as an observable of the
theory if
we insist on having $S^{1}$ as our configuration space.

Alternatively, we can accept $H$ as an observable.  In that case, we must
conclude that quantization drastically changes the topology of the
configuration space from $S^{1}$ to the real line.  Provisionally, we will not
consider such a possibility further in this paper and will insist on having
$S^{1}$ as our configuration space.

Next consider the evolution operator $e^{iT_{0}H}$.  It is single-valued in
$\theta$ [that is, expectation values like ($\Psi _{\theta
_{0}},e^{iT_{0}H}\Psi _{\theta _{0}}$) are single-valued in $\theta _{0}$] if
\be
T_{0} =\frac{1}{c} .\label{2.10}
\ee
With this choice of $T_{0}$, $e^{inT_{0}H}$ are also single-valued if $n\in
\mbox{ the set }Z$ of integers, but not if $0<|n|<1$.  We conclude that time
evolution in this theory is possible only in discrete units of $T_{0}$.
Quantization has quantized time evolution.

Quantization has also altered the nature of energy conservation.  This is
because $H$ is not an observable, and only $e^{iT_{0}H}$ is an observable.
Energy is thus given by a point on a circle and is defined only modulo
$\frac{2\pi}{T_{0}}$.  It is therefore conserved only modulo
$\frac{2\pi}{T_{0}}=2\pi c$.

\sxn{Gauge Transformations and the Nature of Time}

There is another Lagrangian for (\ref{2.1}).  It is time-dependent, but
single-valued and reads
\be
L' =\frac{1}{2}m\dot{\theta}^{2}+ct\dot{\theta}.\label{3.1}
\ee
Its Hamiltonian
\be
H' =\frac{(p-ct)^{2}}{2m}, \; p=m\dot{\theta}+ct \label{3.2}
\ee
is also single-valued and time-dependent.  So $a)$ there is no need for
quantized
time evolution in the quantum physics of (\ref{3.1}), and $b)$ its energy is
not
conserved even modulo $2\pi c$.  Thus although (\ref{2.2}) and
(\ref{3.1}) describe the same classical physics, they do not give the same
quantum physics.

(\ref{2.1}) has a simple physical interpretation: it is the equation of motion
of a particle of unit charge on $S^{1}$ subject to a constant electric field
$E=-c$ tangent to the circle.  This field can be described by either of the
potentials
\begin{eqnarray}
A_{0}=c\theta ,& & A=0  ,  \label{3.3}\\
A'_{0}=0 ,& & A'=-ct .\label{3.4}
\end{eqnarray}
If Lagrangians are constructed in the usual way from these vector
potentials, the former gives (\ref{2.2}) and the latter gives (\ref{3.1}).

Now these potentials are related by a gauge transformation:
\be
A_{\mu}'=A_{\mu} -\frac{\partial}{\partial x^{\mu}}(ct\theta ),\;\;\; x^{0}=t,
\;\;x=\theta . \label{3.5}
\ee
According to conventional understanding, gauge-related potentials describe
identical physics.  But we have seen that such is not the case for $A_{\mu}$
and $A_{\mu}'$.

The reason for this discrepancy lies of course in the multiple-valuedness of
the
function
\be
\Lambda (t,\theta )=ct\theta \label{3.6}
\ee
defining the gauge transformation.  It is precisely because of this property
that the multiple-valued $H$ gets transformed to the single-valued $H'$.

We can now see a general approach to construct multiple-valued Hamiltonians for
any
Lagrangian dynamics on $S^{1}$.  Let $L$ and $H$ be the Lagrangian and
Hamiltonian for some dynamics on
$S^{1}$.  We can assume that they are single-valued.  Consider
\be
L'=L-\frac{d}{dt}(ct\theta ),\label{3.7}
\ee
$\theta$ here being regarded as a function of $t$.
$L'$ and $L$ have the same equations of motion.  If
\[ p=\frac{\partial L}{\partial \dot{\theta}}\;\; ,\;\; H(\theta ,p)
=p\dot{\theta}-L \] are the momentum and Hamiltonian for $L$, then
\begin{eqnarray}
p'=p-ct,\;\; H'=p'\dot{\theta} -L' &=&H[\theta ,p]+c\theta \nonumber \\
&=&H[\theta ,p'+ct]+c\theta \label{3.8}
\end{eqnarray}
are those for $L'$, and $H'$ is multiple-valued because of the potential energy
term
$c\theta$.  Time evolution is quantized in the unit $\frac{1}{c}$ for $H'$.

This construction works in particular for the free particle on $S^{1}$ with the
Hamiltonian, $H=\frac{p^{2}}{2m}$, $p$ being the momentum.

\sxn{Topology and Quantized Evolution}

The first homology group $H_{1}(S^{1},{\bf R})$ of the circle is $Z$ \cite{8}.
There is therefore a closed inexact form $d\theta$ on $S^{1}$.  The existence
of the multiple-valued angle variable $\theta$ on $S^{1}$ can be thought of as
following from this property.

This simple observation suggests that we can hope to find Lagrangians with
quantized time evolution whenever the first homology group $H_{1}(Q,{\bf R})$
of a configuration space $Q$ contains a term $Z$.  Assuming that this condition
is met, let us now show that the above hope is justified.

Because of the stated assumption, there is a closed but inexact one-form $\chi$
on $Q$.  Let $q_{0}$ be a fixed (fiducial) point on $Q$ and $\Gamma _{q}$ a
path from $q_{0}$ to $q$.  Consider
\be
\theta (q)=\int _{\Gamma _{q}}\chi .\label{4.1}
\ee
The right hand side is invariant under deformations of $\Gamma _{q}$ with
$q_{0}$ and $q$ held fixed because of Stokes' theorem \cite{8}, $\chi$ being
closed.  The indication of $\theta$ in (\ref{4.1}) as a function only of $q$ is
therefore also locally justified.

By hypothesis, there are one or more noncontractible closed loops ${\cal
C}_{i}$ generating the homology group $H_{1}(Q,{\bf R})$.  We assume without
loss of generality that the integral of $\chi$ over one such loop ${\cal
C}_{i}$ is $2\pi$ whereas it is zero for any other loop ${\cal C}_{j}$.
There is such a $\chi$ for each choice of ${\cal C}_{i}$.  We now relabel
$\chi$ and $\theta$ as $\chi _{i}$ and $\theta _{i}$ to
indicate this fact.
We then have,
\begin{eqnarray}
\int _{{\cal C}_{j}}\chi _{i} &=&2\pi\delta _{ij} ,\nonumber\\
\theta _{i}(q) &:=& \int_{\Gamma _{q}}\chi _{i}. \label{4.2}
\end{eqnarray}

The function $\theta _{i}$ is a multiple-valued function on $Q$.  It changes by
$2\pi$ when $q$ is taken around (a loop homologous to) ${\cal C}_{i}$ and
behaves like the angular
variable on ${\cal C}_{i}$.

Now given a Lagrangian $\hat{L}_{0}$ for $Q$, we can consider a new Lagrangian
\be
\hat{L}_{i}=\hat{L}_{0} -c\theta _{i}(q),\;\; c>o .\label{4.3}
\ee
It is similar to (\ref{2.2}) and would lead to quantized time evolution with
the time unit $T_{0}=\frac{1}{c}$.

There is also another Lagrangian $\hat{L}'_{i}$ classically equivalent to
$\hat{L}_{i}$.  It is similar to (\ref{3.1}) and reads
\be
\hat{L}'_{i}=\hat{L}_{0} +ct\frac{d\theta _{i}(q)}{dt} .\label{4.4}
\ee
The equations of motion for the Lagrangians $\hat{L}_{i}$ and $\hat{L}_{i}'$
are the same since they differ by
the total derivative
\be
-\frac{d}{dt}[ct\theta _{i}(q)] \label{4.5}
\ee
just as in (\ref{3.7}).  If $\hat{L}_{0}$ is single-valued, then so is
$\hat{L}'_{i}$ and continuous quantum time evolution is possible for
(\ref{4.4}).

The availability of the term (\ref{4.5}) means that we can add it to any
single-valued Lagrangian without affecting its classical dynamics.  It will
then result in only quantized quantum time evolution.

Using (\ref{4.2}), (\ref{4.5}) can be generalised to
\be
-\frac{d}{dt}[t\sum _{i}c_{i}\theta _{i}(q)],\;\; c_{i}>0 .\label{4.6}
\ee
Each term here added to a single-valued Lagrangian gives time unit
$T_{i}=\frac{1}{c_{i}}$.  So if (\ref{4.6}) is added to such a Lagrangian,
then quantum time evolution is possible only if $\frac{T_{i}}{T_{j}}$ is
rational for all $i$ and $j$.  If that is the case, then
$T_{i}=\frac{T}{n_{i}},\;\; n_{i}\in Z^{+}$ and the quantum unit for time
evolution is the least such $T$.

We conclude this section by repeating that quantized evolution can be avoided
by accepting multiple-valued $H$ as observables and accepting as well that
quantization can fundamentally alter topology.  As a working hypothesis, we
have not adopted this point of view in this work.

\sxn{Quantized Evolution in Field Theory}

In field theories in $(N+1)$-dimensional spacetimes, fields at fixed time are
functions on an $N$- dimensional spatial slice with values in a target space
$T$.  The configuration space $Q$ typically consists of such fields.  For
quantized evolution, we must therefore find a $Q$ of this kind with its
fundamental group containing a term $Z$.

It is easy enough to find many models of this kind.  For example let the
spatial slice be $R^{N}$ and let $T$ be the $(N+1)$- dimensional sphere
$S^{N+1}$.  For standard forms of kinetic energy, its finiteness requires that
any field $\xi$ approaches a constant value at spatial infinity for $N>1$,
while we can require $\xi$ to have the same finite value at $\pm \infty$ for
$N=1$.  We can therefore think of $Q$
as maps of $S^{N}$ to $S^{N+1}$ in the usual manner \cite{7}.  The fundamental
group of $Q$ is then given by the homotopy classes of the maps
$S^{N+1}\rightarrow S^{N+1}$ and is $Z$ \cite{7}.  It is also the homology
group
$H_{1}(Q,{\bf R})$.

Another class of examples is provided by $CP^{N}$ models \cite{9}.  The space
$CP^{N}$ is defined as follows.  Let $z_{i}\, (i\leq i\leq N+1)$ be complex
variables constrained by the condition
\be
\sum _{i=1}^{N+1}|z_{i}|^{2}=1 .\label{5.1}
\ee
Then
\be
z\equiv (z_{1},\ldots ,z_{N+1}) \label{5.2}
\ee
are points of a sphere $S^{2N+1}$.  The group $U(1)=\{e^{i\theta}\}$ acts
freely on this space according to the rule
\be
z\rightarrow e^{i\theta}z .\label{5.3}
\ee
Its quotient by this $U(1)$ action is $CP^{N}$:
\be
CP^{N} = S^{2N+1}/U(1) .\label{5.4}
\ee
With ${\bf R}^{1}$ as spatial slice, and with the boundary condition indicated
above at spatial infinity, we see readily from standard results \cite{13} that
$\pi _{1}(Q)=\pi _{2}(CP^{N})=\pi _{1}(U(1))=Z$.

The form of the topological term for $N=1$ is instructive to consider.  As
$CP^{1}$ is the two-sphere $S^{2}$, all formulas can here be written in very
familiar terms.  Note that the preceding sphere models for $T$ also give this
model for $N=1$.

Let us think of $S^{2}$ as the sphere of unit radius centered at the origin in
the Euclidean space ${\bf R}^{3}$.  If $(x,t)$ are the spacetime coordinates,
the field $X=(X^{(1)},X^{(2)}, X^{(3)})$ is then a function of $(x,t)$ subject
to the constraint
\be
\sum _{i}X^{(i)}(x,t)^{2} =1. \label{5.5}
\ee

Our boundary conditions at $\pm \infty$ require that $X(\pm \infty ,t)$ are
equal to the same constant.  We take it to be located at the north pole:
\be
X(\pm\infty ,t)=(0,0,1).\label{5.6}
\ee

The configuration space $Q$ consists of these fields at a fixed time.  A point
$q$
of $Q$ is a field $X$ at a fixed time.  The closed but inexact one-form $\chi$
on $Q$ assigns a real number to any curve in $Q$.  Now suppose that
$\Gamma _{q}$ is a curve
from a fiducial point $q_{0}$ to a point $q$ as required in Section 4.  It is
convenient to take $q_{0}$ to be the constant field with the value (0,0,1).
We can then describe $\Gamma _{q}$ by
\be
\Gamma _{q}=\{ X(\cdot ,t,\tau ):0\leq \tau \leq 1;\; X(x,t,\tau )\in
S^{2}\mbox{ and } X(x,t,0)=(0,0,1)\mbox{
for all }x;\;  X(\cdot ,t,1) =q \} .\label{5.7}
\ee
$\chi$ assigns a real number to $\Gamma _{q}$.

It is easy to see that $\Gamma _{q}$ can be associated with a disc $D$ on the
target $S^{2}$.  Thus for each fixed $\tau$, as $x$ varies from $-\infty$ to
$\infty$, $X(x,t,\tau )$ traces a loop on $S^{2}$ from (0,0,1) to (0,0,1) as in
Fig. 1(a).  As $\tau$ is varied from 0 to 1, this loop sweeps out the disc $D$
mentioned above as in Fig. 1(b).

\begin{figure}[hbt]
\begin{center}\mbox{\psbox{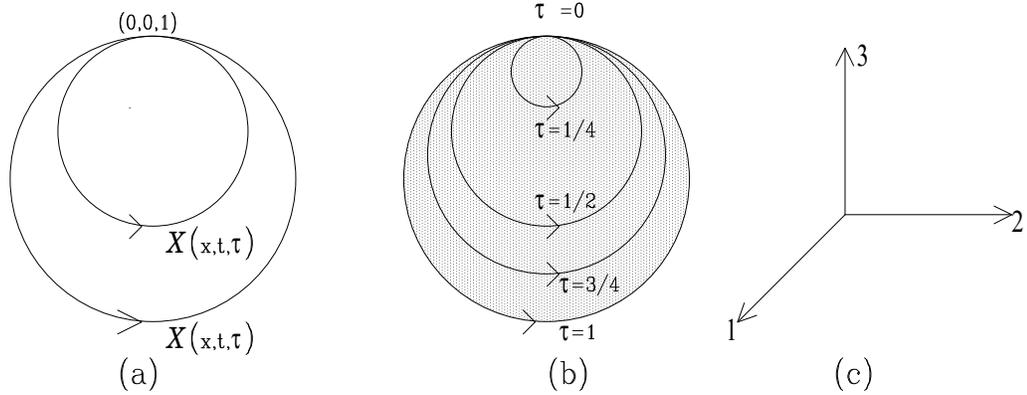}}
\end{center}
\caption{For fixed $\tau$, $X(x,t,\tau)$ traces a loop on $S^{2}$ from
north pole to north pole in Fig.\@ 1(a) when $x$ varies from $-\infty$ to
$+\infty$.  This loop traces out a disk as in Fig.\@ 1(b) when $\tau$ is varied
from 0 to 1.  The arrows show the direction of increasing $x$.  Fig.\@ 1(c)
shows the coordinate system in the embedding space.  The origin of this
coordinate system is the center of the target sphere $T$.}
\end{figure}

Note that
\be
\partial D =\mbox{ Boundary of }D =\{ X(x,t,1):-\infty <x<\infty \}
.\label{5.8}
\ee

It follows that $\chi$ assigns a real number to $D$.

A choice for $\chi$ is given by the volume form on $S^{2}$.  It is defined by
\begin{eqnarray}
\int _{D}\chi &=&\frac{1}{4}\int _{D} \varepsilon _{ijk}X^{(i)}(x,t,\tau)\,
dX^{(j)}(x,t,\tau )\, \wedge dX^{(k)}(x,t,\tau ),   \nonumber\\
&=&\frac{1}{4}\int _{0}^{1}d\tau\int _{-\infty}^{\infty}dx\, \varepsilon
_{ijk}X^{(i)}(x,t,\tau )[\frac{\partial X^{(j)}}{\partial x}(x,t,\tau
)\frac{\partial X^{(k)}}{\partial \tau}(x,t,\tau )-\; j\leftrightarrow
k],\nonumber\\
\varepsilon _{ijk}&=&\mbox{ Levi-Civita symbol with }\varepsilon _{123}=+1.
\label{5.9}
\end{eqnarray}
The choice of normalisation of $\chi$ is dictated by its convenience for
considerations below.

The nontrivial loop on $Q$ which generates $\pi _{1}(Q)$ looks as follows: For
$\tau = 0$, it is at the north pole for all $x$.  Then as $\tau$ increases, it
becomes a bigger and bigger loop until at $\tau =1/2$, it lies in the 2-3
plane.  For $\tau >1/2$, it has $X^{(1)}(x,t,\tau )\leq 0$.
It also steadily shrinks as $\tau$ increases beyond 1/2 becoming the north pole
when $\tau =1$.

\begin{figure}[hbt]
\begin{center}\mbox{\psbox{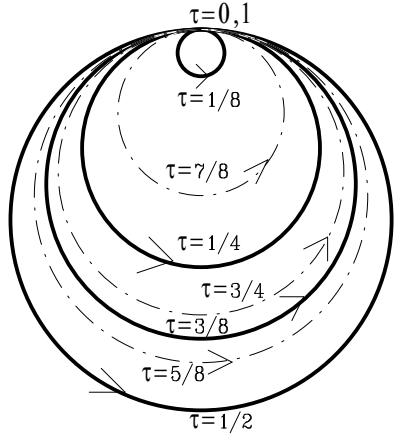}}
\end{center}
\caption{This figure shows a sequence of loops which is itself a
non-trivial loop in $Q$. This loop in $Q$ generates $\pi _{1}(Q)$.  As
explained in the
text, the loops for $\tau >1/2$ are on the far side of the sphere and are
hence shown with dots and dashes while those for $\tau <1/2$ are on the front
side.}
\end{figure}

For this loop, for which $X(x,t,\tau )$ sweeps out the full sphere, (\ref{5.9})
becomes its volume:
\be
\int _{S^{2}}\chi =2\pi .\label{5.10}
\ee

We can now set
\be
\theta (q) =\int _{D}\chi \; ,\; \partial D=\{ X(x,t,1)|-\infty <x<\infty \}
\label{5.11}
\ee
and proceed as before to find Lagrangians which after quantization give only
discrete time evolution.

For completeness, we now write one such specific Lagrangian with $S^{2}$ as
target space, and its equation of motion.  The Lagrangian is
\be
L=\frac{1}{2}\int dx\, \partial _{\mu}X^{(i)}(x,t)\partial
^{\mu}X^{(i)}(x,t) - c\theta (q) \label{5.12}
\ee
where $\theta (q)$ is given by (\ref{5.11}) with $X(x,t,1)=X(x,t)$.
It has the equation of motion
\be
\varepsilon _{ijk}X^{(j)}(\frac{\partial ^{2}}{\partial t^{2}}-\frac{\partial
^{2}}{\partial x^{2}})X^{(k)}+ \frac{c}{2}\frac{\partial}{\partial x}X^{(i)}=0,
\label{5.13}
\ee
which can be obtained from the variation $\delta X^{(i)}=\varepsilon _{ijk}\xi
^{(j)}X^{(k)}$, $\xi ^{(j)}$ being a function of $x,t$ and $\tau$.
\sxn{Quantized Spatial Translations}

There are also models which on quantization admit only discrete spatial
translations even though classical equations of motion have spatial
translational invariance.  We discuss one such model here.

The configuration space for the model in question is
\be
Q=S^{1} \times {\bf R}^{1}=\{ (\theta ,x )\} \label{6.1}
\ee
where $\theta$ and $\theta +2\pi$ are to be identified.  Its Lagrangian is
\be
L= \frac{1}{2I}\dot{\theta}^{2}+\frac{1}{2}m\dot{x}^{2}
+c\dot{\theta}x,\label{6.2}
\ee
$m$, $I$ and $c$ being positive constants.

The classical equations of motion
\begin{eqnarray}
\frac{\ddot{\theta}}{I}+c\dot{x}&=&0,\nonumber\\
m\ddot{x}&=&c\dot{\theta}           \label{6.3}
\end{eqnarray}
are well-defined on $Q$ and are moreover invariant under translations of $x$
and $\theta$ by arbitrary constant amounts.

For quantum theory, (\ref{6.2}) gives the momenta
\be
p= \frac{\dot{\theta}}{I} +cx \; ,\; \pi =m\dot{x}\label{6.4}
\ee
conjugate to $\theta$ and $x$, and the Hamiltonian
\be
H=\frac{I}{2}[p -cx]^{2}+\frac{\pi ^{2}}{2m} .\label{6.5}
\ee

The constant of motion generating translations in $\theta$ is of course $p$.
But as $\pi$ does not commute with $H$,
\be
[\pi ,H]=iIc[p-cx] , \label{6.6}
\ee
it cannot be regarded as the required generator of translations in $x$.  There
is instead another candidate
\be
P=\pi -c\theta \label{6.7}
\ee
for this status since
\be
[P,x]=-i,\;\; [P,H]=0. \label{6.8}
\ee

Note that $P$ does not commute with $p$:
\be
[P,p ] =-ic \label{6.9}
\ee
This is only natural, since $p$ contains the term $cx$ by (\ref{6.4}).

$P$ is thus a good candidate for the generator of spatial translations.  It is
even the unique candidate for this job if we forbid the addition of multiples
of
unity to it.

There is however a serious obstruction to identifying $P$ as this generator: it
is not single-valued just like our previous Hamiltonians.  But the unitary
operator
\be
U=e^{iP/c}\label{6.10}
\ee
is single-valued and hence an acceptable observable.  As $U$ performs spatial
translations in units of $1/c$, we can conclude that spatial translations in
quantum theory can be performed only in units of $1/c$.  This is so even though
classical equations of motion are invariant under continuous spatial
translations.

A final point about this model is as follows.  If the coordinate $x$ as well
describes $S^{1}$, so that $x$ and $x+x_{0}$ must be identified for some
constant $x_{0}$, then it has no quantum time evolution even by discrete
amounts.  One can see this from (\ref{6.5}) which shows that the increment of
$H$ when $x$ changes to $x+x_{0}$ is by the operator $-I(p
-cx)cx_{0}+\frac{I}{2}(cx_{0})^{2}$.  Hence there is no constant $T_{0}$ for
which $e^{iT_{0}H}$ is single-valued in $x$.

Many such models can be constructed using the material in the preceding
sections.
All of them require that the fundamental group of the configuration space
contains the term $Z$.

\sxn{Concluding Remarks: Is Time Itself Quantized?}

We have seen several models where quantum time translations form a discrete
group with a fundamental unit $T_{0}$.  But although these translations are
possible only by discrete steps, it does not quite say that time itself is
discrete.

This point can be made clearer by examining the analogous spatial problem in
Section 6.  Spatial translations came out discrete there in quantum theory.
But space itself remained continuous, the position operator $x$ having a
continuous spectrum.

If time is continuous, and evolution is by discrete jumps, then we are faced
with a conceptual problem which goes as follows.

Let us suppose that we are given the wave function $\Psi (\cdot ,t_{0})$ at
some time $t_{0}$.  Then quantum theory will predict the wave functions $\Psi
(\cdot ,t_{0}+nT_{0})$ only at times $t_{0}+nT_{0}$ where $n\in Z$.

But although theory gives the states only at times $t_{0}+nT_{0}$, experiments
can be performed at all times, time being continuous.  Thus theory has no
prediction for experiments at times $t\neq t_{0}+nT_{0}$.

A way to resolve this problem is to abandon our point of view that $H$ is not
an observable.  In that case, theory does predict $\Psi$ for all times, but
experiments will find that the $\theta$- coordinate of Section 2, for example,
describes a real line and not a circle.

We are thus faced with a difficulty with either interpretation.  A natural
solution, compatible with the point of view of this paper, would therefore be
to assume that time itself is discrete.  Although this suggestion lacks the
compulsive quality of a proof, it strikes us as the most acceptable hypothesis.

We remark in conclusion that we cannot find any reason to suggest spatial
discreteness from the discreteness of spatial translations.  Time and space
after all do not have the same status in physics.

\sxn{Acknowledgements}

We thank the members of our group at Syracuse and colleagues elsewhere, and
especially R.Nityananda, J.Samuel and A.Qamar, for useful and instructive
comments.  This work was supported by the Department of Energy under contract
number DE-FG02-ER40231.

\end{document}